\documentclass[11pt]{article}
\usepackage{epsfig}
\usepackage{graphicx}
\usepackage{amssymb}      
\usepackage{here}  

\evensidemargin=\oddsidemargin  

\newcommand{\beq}{\begin{equation}}
\newcommand{\eeq}{\end{equation}}
\newcommand{\bq}{\begin{equation}}
\newcommand{\eq}{\end{equation}}
\newcommand{\ba}{\begin{array}}
\newcommand{\ea}{\end{array}}
\newcommand{\beqa}{\begin{eqnarray}}
\newcommand{\eeqa}{\end{eqnarray}}

\newcommand{\tr}{{\rm Tr}}
\def\O{{\cal O}}
\def\TT{{\cal T}}

\def\End{\end{document}}
\def\thisday{January, 2002}

\def\to{\rightarrow}

\def\dis{\displaystyle}
\def\f{\frac}
\def\ov{\overline}
\def\[{\left[}
\def\]{\right]}
\def\({\left(}
\def\){\right)}

\def\Ahh{\widehat{A}}
\def\Ah{\widetilde{A}}
\def\pih{\widetilde{\pi}}
\def\At{\widetilde{A}}
\def\pit{\widetilde{\pi}}

\def\l{{\ell}}
\def\tm{{\widetilde{m}}}
\def\Om{{\Omega}}

\def\ct{\cos\theta}

\def\gt{\widetilde{g}}
\def\[{\left[}
\def\]{\right]}
\def\dis{\displaystyle}

\begin{document}
\setcounter{footnote}{1}

\begin{titlepage}
\def\thepage {}        

\title{
\vspace*{7mm}
{\bf Unitarity of Deconstructed Five-Dimensional}\\
{\bf Yang-Mills Theory}
\\[7mm]  }

\author{
{\sc 
R. Sekhar Chivukula%
}\,$^{\rm a}$\footnote{Electronic addresses:
sekhar@bu.edu}  %
~~~and~~~ 
{\sc 
Hong-Jian He%
}\,\,$^{\rm b}$\footnote{Electronic addresses:
hjhe@physics.utexas.edu}%
\\[5mm]
$^{\rm a}$~Department of Physics, Boston University, \\
590 Commonwealth Ave., Boston, Massachusetts  02215, USA \\[2mm]
$^{\rm b}$~Center for Particle Physics,\\ 
University of Texas at Austin, Texas 78712, USA}

\date{\thisday}  

\maketitle

\bigskip
\begin{picture}(0,0)(0,0)
\put(355,295){BUHEP-02-02}
\put(355,280){UTHEP-02-26}
\put(355,265){hep-ph/0201164}
\end{picture}
\vspace{24pt}

\begin{abstract}
\noindent

The low-energy properties of a compactified five-dimensional gauge theory can
be reproduced in a four-dimensional theory with a replicated gauge group and
an appropriate gauge symmetry breaking pattern. The lightest vector bosons in
these ``deconstructed'' or ``remodeled'' theories have masses and couplings
approximately equal to those of the Kaluza-Klein tower of massive vector
states present in a compactified higher-dimensional gauge theory. We analyze
the unitarity of low-energy scattering of the massive vector bosons in a
deconstructed theory, and examine the relationship between the scale of
unitarity violation and the scale of the underlying chiral symmetry breaking
dynamics which breaks the replicated gauge groups. As in the case of
compactified five-dimensional gauge theories, low-energy unitarity is ensured
through an interlacing cancellation among contributions from the tower of
massive vector bosons. We show that the behavior of these scattering
amplitudes is manifest without such intricate cancellations in the scattering
of the would-be Goldstone bosons of the deconstructed theory.  Unlike
compactified five-dimensional gauge theories, the amplitude for longitudinal
vector boson scattering in deconstructed theories does grow with energy,
though this effect is suppressed by $1/(N+1)$, with $N+1$ being the number of
replicated gauge groups.

\pagestyle{empty}
\end{abstract}
\end{titlepage}

\setcounter{section}{0}
\setcounter{equation}{0}
\setcounter{footnote}{0}   


\bigskip

The world may be consistently described by a compactified higher dimensional
theory, manifested via additional towers of massive Kaluza-Klein (KK) states
at low energies.  Recently, it has been shown
\cite{Arkani-Hamed:2001ca,Hill:2000mu} that the low-energy properties of a
compactified five-dimensional gauge theory may be reproduced from a
four-dimensional theory with a replicated gauge group structure and an
appropriate symmetry breaking pattern\footnote{Related issues have been
  considered previously in a variety of contexts, see
  \protect\cite{Eguchi:1982nm,Fu:1984ei,Bando:1988br,Hulsebos:1995pa,Rozali:1997cb,Kachru:1998ys,Bershadsky:1998mb,Sfetsos:1998xd}.}.
A simple scheme is illustrated by the ``aliphatic moose'' shown in Figure\,1.
In this ``moose'' or ``quiver'' diagram \cite{moos,quiver}, the circles
represent $N+1$ $SU(m)$ gauge groups (labeled by $j=0,1,2,\cdots,N$) and the
directed lines represent the Goldstone bosons from the spontaneous symmetry
breaking of the two adjacent $SU(m)$ groups down to their diagonal subgroup.
Thus, we have the gauge symmetry breaking pattern, $SU(m)^{N+1}\to SU(m)_{\rm
  diag}$, generating $N(m^2-1)$ massive spin-1 vector states.

The Goldstone bosons may be collected into SU(m) matrix fields $U_j$
($j=1,2,\cdots,N$) which transform under the adjoining
$SU(m)_{j-1}\otimes SU(m)_{j} $ groups according to
\beq
U_j(x) \to \Om_{j-1}(x) U_j(x) \,\Om^\dagger_j(x)~,
\eeq
where $\Om_{j (j-1)}$ is the $SU(m)_{j (j-1)}$ gauge transformation.
We thus write $U_j$ as \cite{CCWZ},
\beq
U_j(x) = \exp\left({i2\pi^a_j(x)T^a\over v}\right)~,
\eeq
where $\{T^a\}$ are the $SU(m)$ generators (normalized by $\tr \,T^a
T^b = \delta^{ab}/2$), and $v$ is the analog of the QCD pion decay
constant $f_\pi$ which characterizes chiral symmetry breaking.

\begin{figure}[H]
\begin{center}
\includegraphics[width=16cm,height=3cm]{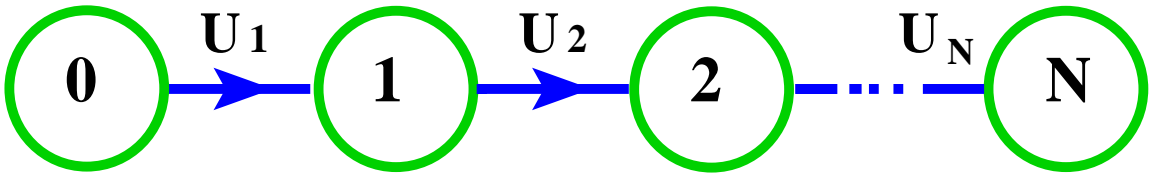} %
\end{center}
\caption{The ``aliphatic moose'' model with $N+1$ replicated 
$SU(m)$ gauge groups.
}
\label{fig:AM}
\end{figure}  

Regardless of the underlying dynamics responsible for the gauge
symmetry breaking, the low-energy properties of this model
may be most economically described by an effective Lagrangian with 
only the gauge and Goldstone degrees of freedom. The leading terms in 
this description are
\beq
{\cal L} = {-\,{1\over 4}} \sum^N_{j=0} F_j^{a\mu\nu} F_{j\mu\nu}^a
+ {v^2\over 4}\sum^N_{j=1} \tr\left(D^\mu U_j D_\mu U^\dagger_j\right)
\,,
\label{eq:L}
\eeq
where the $F_j^{a\mu\nu}$ is the field-strength of gauge group $SU(m)_j$,
and the covariant derivative is,
\beq
D^\mu U_j = \partial^\mu U_j -igA^{a\mu}_{j-1} T^a U_j
                             +igA^{b\mu}_j U_j T^b  \,.
\eeq
Following Ref.\,\cite{Arkani-Hamed:2001ca}, we will refer the above
model as a ``deconstructed'' theory.  In the effective Lagrangian
(\ref{eq:L}), there are $N(m^2-1)$ massive gauge bosons which acquire
their masses from absorbing the corresponding would-be Goldstone
bosons via the Higgs mechanism, and no scalars
remain.  The nonlinear sigma model(s) in the deconstructed theory are
not renormalizable. Naive power counting \cite{Weinberg,GM,georgi}
implies such an effective theory is valid only for scales $\lesssim 4
\pi v$, and the underlying dynamics of chiral symmetry breaking must
become manifest at or below this scale.

Note that we have chosen the gauge couplings ($g$) and the vacuum
expectation values ($v$) of the deconstructed theory to be the same
for all gauge groups and symmetry breakings. This pattern was chosen
so as to reproduce the low-energy properties of a five-dimensional
(5D) $SU(m)$ Yang-Mills theory in which the fifth dimension is
compactified to a line segment $0\le x^5 \le \pi R$. This
compactification can be done consistently by an orbifold projection as
follows: restrict the the gauge fields of the five-dimensional theory
$\Ahh^M(x^N)$ [$M,N\in (\mu,\,5)$ with $\mu \in (0,1,2,3)$] to those
periodic in $x^5$ with period $2\pi R$ and further impose a $Z_2$
symmetry,
\beq
\Ahh^\mu(x^\nu,x^5) = + \Ahh^\mu(x^\nu,-x^5)\,,~~~~
\Ahh^5(x^\nu,x^5)   = - \Ahh^5(x^\nu,-x^5)  \,.
\eeq
These projections force the gauge-covariant boundary conditions,
\beq
\widehat{F}^{5N}=\widehat{F}^{N5}=0
\eeq
at $x^5=0$ and $\pi R$.  Analyzing the KK modes of this compactified
theory shows that, in unitary gauge, one has an infinite tower of
massive $SU(m)$ adjoint vector fields of mass \,$n/R$\,
($n=0,1,\cdots$). In the compactified 5D theory, the self-interactions
of the zero-mode fields are that of a four-dimensional (4D) Yang-Mills
theory with gauge-coupling $g = g^{~}_5 / \sqrt{\pi R}$, where
$g^{~}_5$ is the 5D Yang-Mills coupling with dimension of
(mass)$^{-1/2}$.  The interactions of the KK modes amongst themselves
and with the zero mode gauge-bosons are given by Yang-Mills like
couplings \cite{Hill:2000mu,Dicus:2000hm,Chiv:2001}.

Compactified 5D Yang-Mills theory results in an effective 4D KK theory
which has the remarkable property \cite{Chiv:2001} that low-energy
unitarity is ensured through an interlacing cancellation among
contributions from the relevant KK levels, and is delayed to energy
scales higher than the customary limit of Dicus-Mathur and
Lee-Quigg-Thacker
\cite{Dicus:1973vj,Lee:1977yc,Lee:1977eg,Veltman:1977rt} through the
introduction of additional vector bosons rather than Higgs scalars.
In this Letter, we analyze the unitarity of low-energy massive vector
boson scattering in the deconstructed theory, and examine the
relationship between the scale of unitarity violation and the scale of
the underlying chiral symmetry breaking dynamics responsible for
spontaneously breaking the replicated gauge groups.  We show that
the interactions of the massive vector bosons in the deconstructed
theory are, for levels small compared with $N$, precisely the
same as those of the 4D KK theory. We explicitly show that, 
up to corrections suppressed by $1/N$, the interactions among
the would-be Goldstone bosons encoded in eqn.\,(\ref{eq:L}) match
exactly with the interactions among the corresponding modes of 
$A^{a5}_n$ absorbed through the geometrical Higgs mechanism in the 
compactified 5D gauge theory.

We begin by reviewing the correspondence between the 4D aliphatic
moose model [cf. eqn.\,(\ref{eq:L})] and the orbifold compactification
of 5D Yang-Mills theory \cite{Hill:2000mu}.  Diagonalizing the $N+1$
by $N+1$ dimensional mass-squared matrix in the aliphatic theory, we
find the mass eigenvalues
\beq
M_n = g v \sin{n \pi \over 2(N+1)}~,
\label{eq:masses}
\eeq
with $n\in (0,1,2,\cdots,N)$, which correspond to eigenstates
\beq
\widetilde{A}^a_{0\mu} = {1\over \sqrt{N+1}} (A^a_{0\mu} + A^a_{1\mu} +
\cdots + A^a_{N\mu}) \,,
\label{eq:zeromode}
\eeq
for the remaining massless gauge field, and 
\beq
{\At}^a_{n\mu} = \sqrt{2\over N+1} \sum^N_{k=0}
\cos\left[\left(k+{1\over 2}\right)\,{n\pi \over N+1}\right] 
A_{k\mu}^a \,,
~~~~~~ (n=1,2,\cdots,N)\,,
\label{eq:nonzeromodes}
\eeq
for the the massive adjoint vector bosons.  The massless fields
$\{\At^a_{0\mu}\}$ belong to the residual unbroken gauge group
$SU(m)_{\rm diag}$ with coupling 
~$\widetilde{g}=g/\sqrt{N+1}$\,.

Comparing the mass spectrum (\ref{eq:masses}) of the deconstructed theory
(for $n \ll N+1$) with the linear spectrum $n/R$ in the KK theory, 
we see that the two coincide under the identification 
\cite{Arkani-Hamed:2001ca,Hill:2000mu},
\beq
{1\over R} = {\pi g v \over \,2(N+1)\,}
           = {\pi \,\gt\, v \over \,2\sqrt{N+1}\,}  \,,
\label{eq:R}
\eeq
and eqn.\,(\ref{eq:masses}) can be expanded as,
\beq
M_n =\dis\f{n}{R}\[1-\f{\pi^2}{24}
         \(\f{n}{N+1}\)^2 +O\!\(n^4\over N^4\)\]
\equiv \ov{M}_n\[1-\delta_n\] ,
\label{eq:massApp}
\eeq
where \,$\ov{M}_n=n/R$\, and $\,\delta_n = \O(n^2/N^2)\,$.
The identification (\ref{eq:R}) 
corresponds to interpreting the Moose diagram itself
as a discretized fifth-dimension with a lattice spacing,
\beq
\dis
a = {\,\pi R\, \over N} = {2 \( 1+N^{-1} \)\over g v} 
                   = \f{\,2\sqrt{1+N^{-1}}\,}{\gt\,v\sqrt{N}} \,.
\label{eq:latticespacing}
\eeq
The spectrum of the deconstructed theory approximates the 
linear spectrum of the 4D KK theory so long as \,$n \ll N+1$,\,
i.e., \,$M_n \ll 1/a$\,.

The correspondence between the deconstructed theory and
4D KK theory may be completed by identifying the couplings
of the unbroken massless gauge group,
\beq
{1\over \,{\gt}^2\,}  = {\,\pi R\, \over g^2_5}  ~,
\label{eq:Rg5gt}
\eeq
which yields
\beq
g^{~}_5 = \sqrt{\,2 g\, \over v}~.
\label{eq:g5value}
\eeq
This correspondence implies that
\beq
{a \over \,g^2_5\,} = \f{1}{\,N\,\gt^2\,} 
= {\,1+N^{-1}\, \over g^2}~,
\eeq
and the ``bare'' coupling of the deconstructed theory, 
$g$, may be identified with the effective strength of 
the gauge coupling in the 5D Yang-Mills theory 
underlying the 4D KK theory.

As indicated in eqn.\,(\ref{eq:massApp}), the exact correspondence between the
spectra of the deconstructed theory and the 4D KK theory is realized only for
sufficiently large $N$.   We will explicitly show that the vector-boson
scattering amplitudes agree in these theories as well, up to corrections
suppressed by $1/N$.  The ``continuum limit'' corresponds to $a \to 0$ for
fixed  $R$ and $\gt$, or, equivalently,  
  $N\to \infty$ with\footnote{As 
  a practical matter, of course, 
  the coupling of the replicated gauge groups $g$ is bounded by
  ${\cal O}(4\pi)$.  Hence, there is a bound on how large $N$ can be for
  a fixed size of the low-energy coupling $\gt$\,. This is similar to the bound
  on the underlying scale of the 5D gauge theory relative to the compactification
  scale arising from the constraint that the 4D gauge coupling has a
  finite size.} \,$g = {\cal O}(\sqrt{N}\,)$\,.  
From eqn.\,(\ref{eq:R}),
therefore, we deduce that $v = {\cal O}(\sqrt{N}\,)$ 
when approaching the continuum limit.

A 5D gauge theory is nonrenormalizable, and one manifestation of this
is the bad high-energy behavior of massive vector-boson scattering.  In
Ref.\,\cite{Chiv:2001}, it is shown that tree-level gauge boson scattering in
the 5D $SU(m)$ Yang-Mills theory violates unitarity at an energy scale of the
order
\beq
\sqrt{s} ~=~ E_{\rm cm} ~\leq~ \Lambda 
          = {96 \pi \over \,23\, m\,}\,{1\over g^2_5}  \,,
\label{eq:Ubound5D}
\eeq
and therefore this theory is, at best, a low-energy effective theory valid
only up to a scale of order $\Lambda$\,.  In the deconstructed theory, from
eqn.\,(\ref{eq:Rg5gt}),  $\Lambda$ corresponds to an energy scale of order
\beq
\Lambda \simeq {96 \pi \over \,23\,m\,} \,
 \f{1}{\,\gt^2\,N\,a\,} \, ,
\eeq
which is {\it higher} than \,$1/a$\, so long as
\beq
\gt \lesssim {3.6 \over \sqrt{m\,N}~}\, .
\eeq
When the deconstructed theory is embedded into a 4D renormalizable
high-energy theory \cite{Arkani-Hamed:2001ca,Hill:2000mu}, the 4D theory
provides a ``high-energy'' completion of the compactified 5D Yang-Mills
theory.  From the considerations above, we see that for weak or moderate
coupling and modest $N$, a deconstructed theory provides a high-energy
completion which respects the bound in eqn.\,(\ref{eq:Ubound5D}).

As noted above, the deconstructed theory itself involves chiral
symmetry breaking dynamics in order to provide the Goldstone-boson
``link'' fields that allow particles to ``hop'' in the fifth
dimension.  Power counting \cite{Weinberg,GM,georgi} shows that the
non-linear sigma model low-energy description must break down at a
scale $\lesssim 4 \pi v$\,.  Given the effective lattice spacing
eqn.\,(\ref{eq:latticespacing}), we see that, $1/a < 4 \pi v$,
provided
\beq
\gt ~\lesssim~ 8\pi \,{\sqrt{N+1}\over N} \,.
\eeq
In this case the non-linear sigma model description can remain valid up to
the scale $1/a$, at which the model no longer behaves like the compactified
effective 4D KK theory.  
In what follows we will investigate the scattering of massive
vector-bosons and their corresponding Goldstone bosons in the deconstructed
theory for energy scales {\it less} than $1/a$, therefore we need not be
concerned about the underlying 4D chiral symmetry breaking dynamics.

To analyze the relevant scattering processes, we start by deriving the 
unitary gauge Lagrangian of the deconstructed theory in Fig.\,1.
In this gauge all link-fields $\{U_j\}$ are set 
to the identity via appropriate $SU(m)$ gauge transformations.
Expressing all the vertices in terms of the mass-eigenstate
gauge fields, we derive the interaction Lagrangian
\beqa
{\cal L}_{\rm gauge} & = & 
- \,\gt C^{abc} \sum_{n=1}^N
\[\partial_{\mu} \At^{a}_{0\nu} \At^{b \mu}_n \At^{c \nu}_n + 
\partial_{\mu} \At^{a}_{n\nu} (\At^{b\mu}_0 \At^{c\nu}_n + 
               \At^{b\mu}_n\At^{c\nu}_0)\]
\nonumber 
\\
&& - {\gt\over \sqrt{2}} C^{abc} \sum_{n,m,l=1}^N 
\Delta_3(n,m,\l) \,
\partial_{\mu} \At^{a}_{n\nu}
\At^{b\mu}_m \At^{c\nu}_\l 
\nonumber 
\\
&&- {\gt^2\over 4} C^{abc} C^{ade} \sum_{n=1}^N
\[\At^{b}_{0\mu} \At^{c}_{0\nu} \At^{d\mu}_n \At^{e\nu}_n
+ {\rm all\ permutations}\] 
\label{eq:selfint}
\\  
&& -{\gt^2 \over 4\sqrt{2}} C^{abc} C^{ade} \sum_{n,m,\l =1}^N 
\Delta_3(n,m,\l) 
\[\At^{b}_{0\mu} \At^{c}_{n\nu} \At^{d\mu}_m \At^{e\nu}_\l 
+ {\rm all\ permutations}\] 
\nonumber 
\\
&& 
-{\gt^2\over 8} C^{abc} C^{ade} \sum_{n,m,\l,k=1}^N 
\Delta_4(n,m,\l,k) \,
\At^{b}_{n\mu} \At^{c}_{m\nu} \At^{d\mu}_\l \At^{e \nu}_k  ~, \nonumber
\eeqa
with $\Delta_3$ and $\Delta_4$ given by
\beqa
\Delta_3 (n,m,\l)
& \!=\! & \delta(n+m-\l)+\delta(n-m+\l)+\delta(n-m-\l) \,,
\nonumber \\
\Delta_4 (n,m,\l,k)
& \!=\! & \delta(n+m+\l-k)+\delta(n+m-\l+k)+\delta(n-m+\l+k)+ 
\label{deltas} \\
&& 
    \delta(n+m-\l-k)+\delta(n-m-\l+k)+\delta(n-m+\l-k)+\delta(n-m-\l-k) \,, ~~~~~
\nonumber
\eeqa
so long as\footnote{If $(n,\,m,\,\l,\,k)$ are such that $n+m+\l$ or
  $n+m+\l+k$ equals $2q(N+1)$ for $q=1,2,\cdots$, the factors
  $\Delta_{3,4}$ will have an additional contribution equal to
  $(-1)^q$.}  $\,(n,\,m,\,\l,\,k) \ll N+1$\,.  These interactions are
{\it precisely} those found in the 4D KK theory
\cite{Chiv:2001,Dicus:2000hm,Hill:2000mu}.

The deconstructed theory describes a set of massive self-interacting vector
bosons with a characteristic coupling \,$\gt$\,.  The traditional arguments
\cite{Dicus:1973vj,Lee:1977yc,Lee:1977eg,Veltman:1977rt} suggest that the
scattering amplitudes of longitudinally polarized vector bosons at level
$n\ll N+1$ would grow with energy and violate unitarity at an energy scale,
\beq
\dis
E^\star \sim {\,4\pi M_n\over \gt\,} 
\approx  
  \f{2\pi^2 n\,v}{\,\sqrt{N+1}\,\,} 
= \f{\,4n\pi^2\,}{\,\gt \,N \,a\,}
= \f{\,4n\pi^2 \,\gt\,}{g_5^2}  
~.
\label{eq:UB0}
\eeq
This cannot be the case because, for given $v$, $E^\star$ could be made much
smaller than $4\pi v$ --  the scale at which the chiral symmetry breaking
dynamics is expected to enter the deconstructed theory.

Furthermore, $E^\star$ could be made much smaller than the scale $1/a$ below
which the theory should correspond to the 4D KK theory.  This provides a clue
for how to proceed: the vector boson masses in the 4D KK theory arise through
a geometrical Higgs mechanism, and unitarity is ensured through an
interlacing cancellation among the contributions of the relevant gauge KK
modes \cite{Chiv:2001}.  We therefore expect that there will be similar
cancellations in the deconstructed theory.  Since the unitary-gauge
Lagrangian (\ref{eq:selfint}) is identical to that derived in the
compactified 5D theory \cite{Chiv:2001,Dicus:2000hm,Hill:2000mu} for low
KK-levels, the deviations of the vector-boson amplitude in the deconstructed
theory from that in the compactified 5D theory can only come from the
modification term in the mass-spectrum (\ref{eq:massApp}).  After a careful
analysis of the gauge amplitudes 
with the mass expansion of (\ref{eq:massApp})
and the high energy expansion of $M_n/E$, 
we find that the individual $\O(E^4)$
terms completely cancel and the non-canceled $\O(E^2)$ terms appear
only at the order $1/N$.  
For instance, we derive the following $\O(E^2)$
amplitudes for the elastic scattering of
the longitudinal gauge bosons,
$\Ah^{an}_L \Ah^{bn}_L \!\to\! \Ah^{cn}_L \Ah^{dn}_L$\,,
\bq
\!{\cal T}\!\[\Ah^{an}_L \Ah^{bn}_L \!\to\! \Ah^{cn}_L \Ah^{dn}_L\]
\!=\!\dis \f{g^2\delta_n\,s}{\ov{M}_n^2(N+1)}  
\!\[
-c\, C^{abe}C^{cde} + \(\f{9}{2}\!+\!\f{11}{2}c\)\! C^{ace}C^{bde}
                    + \(\f{9}{2}\!-\!\f{11}{2}c\)\! C^{ade}C^{bce}
\],
%
\label{eq:ALnn-nn}
\eq
which, as expected, is proportional to the nonlinear modification
$\delta_n$ in eqn.\,(\ref{eq:massApp}), 
and has the coefficient 
$\(g^2\delta_n\)\!/[\ov{M}_n^2(N+1)] \simeq 1/[6v^2(N+1)]$
suppressed by $1/(N+1)$.
Here, $c=\ct$ and $\theta$ is the scattering angle.
For the inelastic scattering 
$\Ah^{an}_L \Ah^{bn}_L \!\to\! \Ah^{cm}_L \Ah^{dm}_L$
($n\neq m$), we arrive at
\bq
%
\!{\cal T}\!\[\Ah^{an}_L \Ah^{bn}_L \!\to\! \Ah^{cm}_L \Ah^{dm}_L\]
=\!\dis \f{s}{6v^2(N+1)}  
\[
-2c\, C^{abe}C^{cde} 
+ \(3\!+\! 5c\) C^{ace}C^{bde}
+ \(3\!-\! 5c\) C^{ade}C^{bce}
\], 
%
\label{eq:ALnn-mm}
\eq
which, with the aid of Jacobi identity
$C^{abe}C^{cde} + C^{ace}C^{dbe} + C^{ade}C^{bce} = 0 $,\,
can be related to the $\O(E^2)$ elastic amplitude 
(\ref{eq:ALnn-nn}) via
\bq
{\cal T}\!\[\Ah^{an}_L \Ah^{bn}_L \!\to\! \Ah^{cm}_L \Ah^{dm}_L\]
\simeq 
\dis\f{2}{3}
{\cal T}\!\[\Ah^{an}_L \Ah^{bn}_L \!\to\! \Ah^{cn}_L \Ah^{dn}_L\]
\,.
\label{eq:nnnn-nnmm}
\eq

The unitarity analysis above is performed in the vector boson sector of the
deconstructed theory.  Similar high energy behavior must also arise in the
would-be Goldstone boson sector of the deconstructed theory.  Since the
deconstructed theory is based on the spontaneous symmetry breaking
$SU(m)^{N+1}\to SU(m)_{\rm diag}$, in the corresponding $R_\xi$ gauge
gauge [cf. eqn.\,(\ref{eq:Lgf})] we can derive the equivalence theorem
\beq
\TT\[\Ah^{a}_{nL}(p_n), \Ah^{b}_{mL}(p_m), \cdots\]  = C_{\rm mod}\,
\TT\[\pih^{a}_n(p_n), \pih^{b}_m(p_m), \cdots\]  + {\cal O}(M_{n,m,\cdots}/E) \,,
\label{eq:DC-ET}
\eeq
where the levels $(n,m,\cdots) = 1,2,\cdots,N$, and each external momentum is
put on mass-shell, $p_n^2=M_n^2$, etc.  Here the fields $\{\pih^a_n\}$ are
the would-be Goldstone bosons ``eaten'' by the corresponding mass eigenstate
vector fields $\{\At^a_n\}$.  This is analogous to the traditional
equivalence theorem in the Standard Model
\cite{Cornwall:1974km,Lee:1977eg,Chanowitz:1985hj,
  Yao:1988aj,Bagger:1990fc,He:1992ng,He:1994yd,He:1997cm}, and the
modification factor $\, C_{\rm mod}=1+\O({\rm loop})\,$ appears at loop level
\cite{Yao:1988aj,Bagger:1990fc,He:1992ng,He:1994yd,He:1997cm}.

To analyze the Goldstone boson scattering, 
we will derive the complete $R_\xi$ gauge Lagrangian 
for the deconstructed theory. 
From the nonlinear dimension-2  term of Goldstone boson kinetic energy in
eqn.\,(\ref{eq:L}), 
we deduce the following bilinear gauge-Goldstone mixings,
\beq
{\cal L}_{\rm mix} =  \dis
\sum^N_{j=1} -\f{1}{2}gv\[A_{j-1}^{a\mu}\partial_\mu\pi^a_j
                     -A_j^{a\mu}\partial_\mu\pi_j^a\]
= \sum^N_{n=1} -M_n\Ah_n^{a\mu}\partial_\mu\pih_n^a \,.
\label{eq:A-GB}
\eeq
Here $M_n$ is given by eqn.\,(\ref{eq:masses}) and the fields
$\{\pih^a_n\}$ are the eigenstates of ``eaten'' Goldstone bosons
defined by the orthogonal rotation,
\beq
\pih^a_n ~=~ \dis\sum^N_{k=1} \sqrt{\f{2}{N+1}}
             \sin\!\(\f{nk\pi}{N+1}\)
             \pi^a_k \,, ~~~~~~(n=1,2,\cdots ,N)\,.
\label{eq:eaten-GB}
\eeq
The bilinear mixing (\ref{eq:A-GB}) can be eliminated by defining a
general $R_\xi$ gauge fixing term,
\beqa
{\cal L}_{\rm GF} &=& - \sum_{n=0}^N {1 \over \,2 \,\xi_n} 
\( \partial^\mu \Ah^{a}_{n\mu}
 - \xi_n M_n \pih^{a}_n  \)^2~,     
\label{eq:Lgf}
\eeqa
where $n=0$ corresponds to the usual gauge-fixing of the unbroken group
$SU(m)_{\rm diag}$. 
The would-be Goldstone bosons $\pih^a_n$ acquire
gauge-dependent masses $\,M_{\pih_n}^2 = \xi_n M_n^2\,$. The
appropriate ghost Lagrangian can be derived as well, though it is not
explicitly needed for the current analysis.

The interactions of the Goldstone bosons with the gauge bosons and among
themselves arise from the nonlinear sigma model, \`{a} la
Callan-Coleman-Wess-Zumino (CCWZ) \cite{CCWZ}. However, the ``geometric''
Goldstone sector in compactificted 5D Yang-Mills theory appears very
different since its Goldstone bosons $\{A^{a5}_n \}$, the fifth components
of the 5D gauge fields, interact at most bi-linearly with other gauge modes,
and have no self-interaction among themselves \cite{Chiv:2001}. How does this
highly nonlinear CCWZ Goldstone sector match with the geometric, linearized
Goldstone sector in compactificted 5D gauge theory?  As we now show, the
correspondence between the Goldstone sector of the deconstructed theory and
that of the compactified 5D Yang-Mills theory works at $\O(N^0)$, with the 
corrections suppressed by powers of $1/N$.

Using the orthogonal rotations (\ref{eq:nonzeromodes})
and (\ref{eq:eaten-GB}), and collecting the terms of order
$N^0$ and $N^{-1}$,
we arrive at, after a lengthy derivation,
the following Goldstone interactions,
\beqa
{\cal L}^{\rm LO}_{\rm GB} & = & 
+ \,\gt C^{abc} \sum_{n=1}^N 
\Ah^{b\mu}_0 \,\pih^{c}_n 
\(\partial^\mu \pih^{a}_n - M_n \Ah^{a\mu}_n\)
+ {\gt^2 \over 2} C^{abc} C^{ade} \sum_{n=1}^N 
\Ah^{b}_{0\mu}\, \Ah^{d\mu}_0 \,\pih^{c}_n \,\pih^{e}_n 
\nonumber   
\\
&&
+ {\gt \over \sqrt{2}} C^{abc} \sum_{n,\tm ,\l=1}^N
\widetilde{\Delta}_3(n,\tm,\l) \,
\Ah^{b}_{n\mu} \,\pih^{c}_{\tm}
\(\partial^\mu \pih^{a}_\l - M_\l \Ah^{a\mu}_\l \)
\nonumber 
\\[-2mm]
&& \label{eq:LGB-N0}
\\[-2mm]
&& 
+{\gt^2\over \sqrt{2}} C^{abc} C^{ade} \sum_{n,\tm ,\l=1}^N
 \widetilde{\Delta}_3(n,\tm ,\l) \,
\Ah^{b}_{0\mu} \Ah^{d\mu}_n \,\pih^{c}_{\tm} \,\pih^{e}_\l 
\nonumber 
\\
&& +{\gt^2\over 4} C^{abc} C^{ade} \sum_{n,\tm ,\l,k=1}^N 
\widetilde{\Delta}_4(n,\tm ,\l,k) \,
\Ah^{b}_{n\mu}\, \Ah^{d\mu}_{\tm} \,\pih^{c}_\l \,\pih^{e}_k 
\,,
\nonumber 
\\[2mm]
{\cal L}^{\rm NLO}_{\rm GB} 
& = & 
\dis\!\!\!\!\sum_{n,\tm ,\l,k=1}^N\!\!
\f{1}{~12(N+1)v^2~}
\widetilde{\Delta}_{4\pi}(n,\tm ,\l,k)
\[\f{2}{m}\delta^{ab}\delta^{cd} + d^{abe}d^{cde}
\]
\nonumber 
\\[2mm]
& &
\dis
\hspace*{10mm}
\times\[ 
\,\pit^a_n\, \partial^\mu \pit^b_{\tm}\, 
  \pit^c_\l\,\partial_\mu \pit^d_k
-
\,\partial^\mu\pit^a_n\,\partial_\mu\pit^b_{\tm}
  \(\pit^c_\l\,\pit^d_k\)
\] +\O\(\gt\),
\label{eq:LGB-N1}
\eeqa
where
\beqa
\widetilde{\Delta}_3(n,\tm ,\l) \!\!\!&\! = &\!\!
\,\delta(n+\tm-\l) + \delta(n-\tm+\l) - \delta(n-\tm-\l)\,,
\nonumber 
\\
\widetilde{\Delta}_4(n,\tm,\l,k) \!\!\!&\! = &\!\! 
\,\delta(n+\tm+\l-k) +\delta(n+\tm-\l+k)
+\delta(n-\tm+\l+k) + \delta(n-\tm-\l+k)~~~~~~~
\nonumber 
\\
\!\!\!&&\!\!\!
\hspace*{-2mm}
-\delta(n+\tm-\l-k) - \delta(n-\tm+\l+k) - \delta(n-\tm-\l-k) 
\,,
\\
\!\widetilde{\Delta}_{4\pi}(n,\tm,\l,k) \!\!\!&\! = &\!\! 
\,\delta(n+\tm+\l-k) +\delta(n+\tm-\l+k)
+\delta(n+\tm-\l-k) 
\nonumber 
\\
\!\!\!&&\!\!\!
\hspace*{-2mm}
-\delta(n+\tm-\l+k) - \delta(n+\tm+\l-k) 
-\delta(n-\tm+\l+k) - \delta(n-\tm-\l-k)
\,. \nonumber
\eeqa
In (\ref{eq:LGB-N1}), the factor $\f{2}{m}$ contains the $m$ of
$SU(m)$; and the symmetric $d$-function is defined by ~$\{T^a,\,T^b\}
=\f{1}{m}\delta^{ab}+d^{abc}T^c$~. The unspecified terms of $\O(\gt)$
or smaller in (\ref{eq:LGB-N1}) contain at most one partial
derivative and at least one gauge field, 
and are irrelevant to the $\O(E^2)$ leading behavior of
the Goldstone scattering amplitude to be derived shortly. 
Other contributions suppressed by $1/N^2$ or higher will also 
not be needed below.

It is important to note that the leading order (LO) Goldstone Lagrangian
${\cal L}^{\rm LO}_{\rm GB}$ in (\ref{eq:LGB-N0}) contains at most two
Goldstone fields. It {\it precisely} matches\footnote{At the leading order of
  the $1/N$-expansion the mass spectrum of gauge fields $\{\Ah^{a\mu}_n\}$
  also becomes identical to that of the the compactified 5D Yang-Mills
  theory, i.e., $M_n = (n/R)\,[1 + \O(n^2/N^2)]\,$,\, as shown in
  eqn.\,(\ref{eq:massApp}).}  with that derived in the compactified 5D
Yang-Mills theory \cite{Chiv:2001}, under the identification of \,\,$\pih^a_n
\longleftrightarrow A^{a5}_{n}$\,.\, To order $N^0$, the correspondence
between the CCWZ Goldstone sector of the deconstructed theory and the
geometric, linearized Goldstone sector in the compactificted 5D gauge theory
is exact. At this order all non-renormalizable interaction vertices
containing of dimension $>4$ disappear, and the leading order
high-energy behavior of Goldstone boson scattering matches that of
compactified 5D Yang-Mills theory \cite{Chiv:2001}.

The deviation of the deconstructed Goldstone Lagrangian from that of the
compactified 5D gauge theory explicitly appears at 
the next-to-leading order (NLO) of the $1/N$-expansion. The dimension-6
quartic Goldstone vertices in (\ref{eq:LGB-N1}) contain two partial
derivatives, analogous to the usual chiral Lagrangian of low energy QCD
\cite{Weinberg}.  The additional factor $1/(N+1)$ in
(\ref{eq:LGB-N1}) indicates that the interactions, and therefore the
amplitudes of scattering, among the eigenstate would-be Goldstone bosons
are suppressed by $N+1$.  
For the scattering processes $\pih^a_n \,\pih^b_n
\to \pih^a_n \,\pih^b_n$ and $\pih^a_n \,\pih^b_n \to \pih^a_m \,\pih^b_m$
($n\neq m$), at $\O(1/N)$ and $\O(E^2/v^2)$,
we derive
\bq
\ba{l}
\hspace*{-6mm}
{\cal T}\[\pih^a_n \,\pih^b_n \to  \pih^a_n \,\pih^b_n\]
=~ \dis 
\f{3}{2}{\cal T}\[\pih^a_n \,\pih^b_n \to  \pih^a_m \,\pih^b_m\]
= \dis \f{3}{\,2(N+1)v^2\,}
\[
s\,X^{ab,cd} + 
t\,X^{ac,bd} +
u\,X^{ad,bc}
\], 
\nonumber
\ea
\label{eq:GB2-2}
\eq
where 
$\, X^{ab,cd} \equiv \f{2}{m}\delta^{ab}\delta^{cd} + d^{abe}d^{cde} $.\, For
$SU(m)=SU(2)$, $X^{ab,cd} = \delta^{ab}\delta^{cd} $ and (\ref{eq:GB2-2})
reduces to the familiar form of the $\pi\pi$ scattering of low energy QCD
\cite{Weinberg} except an overall factor $\sim\! 1/(N+1)$.  
Making use of the $SU(m)$ identity 
\bq 
\dis
C^{abe}C^{cde} ~=~  
\f{2}{m}
\(\delta^{ac}\delta^{bd}-\delta^{ad}\delta^{bc}\)+
\(d^{ace}d^{bde}-d^{ade}d^{bce}\) \,,
\eq
we find that the gauge amplitudes (\ref{eq:ALnn-nn})-(\ref{eq:ALnn-mm})
fully agree to the corresponding Goldstone amplitudes
(\ref{eq:GB2-2}) at the same order of $1/N$, satisfying the
equivalence theorem (\ref{eq:DC-ET}). 

Projecting the elastic amplitude ${\cal T}\[\pih^a_n \,\pih^b_n \to
\pih^a_n \,\pih^b_n\]$ to the isospin-singlet and spin-0 channel,
i.e., \,$\TT_{\,00}^{~}[nn;nn] = [3/2(N+1)][s/16\pi^2v^2]$\, for
$SU(2)$, we readily derive the unitarity bound,\footnote{ Here we do
  not include the contribution of the leading order Lagrangian ${\cal
    L}^{\rm LO}_{\rm GB} $ to the scattering amplitude since it
  behaves as constant and does not grow with the energy, as computed
  in Ref.\,\cite{Chiv:2001}.}
\beq
\dis
\TT_{\,00}^{~}[nn;nn] \,~\leq\,~ \f{1}{2} \,, 
~~~~~~\Longrightarrow~~~~~~
\sqrt{s} ~\leq~ \sqrt{N+1}\,\f{~4\pi v~}{\,\sqrt{3\pi}\,} \,,
\label{eq:UB-delay}
\eeq
which is, apparently, delayed relative to the customary unitary limit
for $\pi\pi$ scattering by a factor of $\sqrt{(N+1)(2/3)}$\,.

However, the deconstructed theory has many ``KK'' levels of $\pih^a_n$, with
$n=1,2,\cdots N$, and we must consider coupled channels as well.  Consider a
normalized state, consisting of Goldstone boson pairs with ``KK'' levels up
to $N_0$\,,
\beq
\left|\Psi^{ab}\right\rangle = 
{1 \over \sqrt{N_0}} \sum_{\l=1}^{N_0} 
\left| \pih^{a}_\l \pih^{b}_\l \right\rangle \,,
\eeq
from which we deduce the $\O(1/N)$ scattering amplitude, at high energies
$\sqrt{s} \gg 2M_{N_0}$\,,
\bq
\ba{l}
\hspace*{-5mm}
\TT \[ \left|\Psi^{ab}\right\rangle 
\to \left|\Psi^{cd}\right\rangle \] 
= \dis \sum_{\l,k=1}^{N_0} 
\f{1}{N_0} {\cal T}\[\pih^a_\l \,\pih^b_\l \to  \pih^a_k \,\pih^b_k\]
\\[5mm]
\hspace*{-3mm}
~= \dis\left.
(N_0-1) {\cal T}\[\pih^a_n \,\pih^b_n \to  \pih^a_m \,\pih^b_m\]\right|_{n\neq m}
+ {\cal T}\[\pih^a_n \,\pih^b_n \to  \pih^a_n \,\pih^b_n\]
\\[4mm]
\hspace*{-3mm}
~=\(N_0+\f{1}{2}\) {\cal T}\[\pih^a_n \,\pih^b_n \to  \pih^a_m \,\pih^b_m\]
%
=\dis \f{(N_0+\f{1}{2} )}{\,(N+1)v^2\,}
\[
s\,X^{ab,cd} + 
t\,X^{ac,bd} +
u\,X^{ad,bc}
\].
\\[2.5mm]
\ea
\eq
Thus, we see that when the number of invoked ``KK'' levels reaches
$N_0\sim N$, we recover the customary unitarity limit, 
for $SU(m)=SU(2)$\,,\footnote{The limit for the arbitrary $SU(m)$ 
varies only by a factor of $\O(m/2)=\O(1)$, 
with no conceptual difference.}
\beqa
\sqrt{s} & \lesssim & \sqrt{8\pi} v 
           = \sqrt{2\over\pi}\,\f{~4g~}{g_5^2}
           = \sqrt{\f{2}{\pi}}\f{\,4(1+N^{-1})\,}{g\,a} \,, 
\eeqa
which is of the order \,$v \sim g/g_5^2 \sim 1/(ga)$,\, and is  
neither enhanced by extra $\sqrt{N+1}$ [cf. the
single channel analysis in eqn.\,(\ref{eq:UB-delay})] 
nor further suppressed by $1/\sqrt{N+1}$ 
[cf. the naive estimate in eqn.\,(\ref{eq:UB0})].

In summary, we have systematically analyzed the gauge and Goldstone
interaction Lagrangians in four-dimensional deconstructed Yang-Mills theory.
For the low ``KK'' levels, the gauge sector differs from the compactified 5D
theory only in the mass-spectrum, but the Goldstone sector explicitly differs
in its interaction Lagrangian at order $\O(1/N)$.  We have analyzed the
relationship between the scale of unitarity violation in longitudinal
vector-boson scattering and the scale of the underlying chiral symmetry
breaking dynamics responsible for spontaneously breaking the replicated gauge
groups.  As in compactified 5D gauge theory, the low-energy unitarity of
longitudinal vector-boson scattering is ensured through an interlacing
cancellation among contributions from various ``KK'' levels. We have shown
that the behavior of these amplitudes can be also understood in the
deconstructed theory by analyzing would-be Goldstone boson scattering via the
equivalence theorem.  Taking into account the non-cancelled
$E^2$-contributions at the order $1/N$, we find that unitarity violation in
the deconstructed theory is delayed to the intrinsic ultraviolet scale
$1/g_5^2$ or $1/a$, and is above the customary Dicus-Mathur/Lee-Quigg-Thacker
limit.  We have also demonstrated explicitly the correspondence between the
Higgs mechanism in the 4D deconstructed theory and the ``geometric Higgs
mechanism'' in the compactified 5D gauge theory.


\vspace*{5mm}
\noindent
{\large {\bf Acknowledgments}}~~~  \\[2mm] 
We thank Duane Dicus, Howard Georgi and Chris Hill for discussions.
This work was supported by the Department of
Energy under grants DE-FG02-91ER40676 and DE-FG03-93ER40757.


\bibliography{deconstruct.bib}
\bibliographystyle{h-physrev3}

\end{document}